%% file: cied_dilepton_7tev.tex
\documentclass[aps,prd,twocolumn,showpacs,nofootinbib,floatfix]{revtex4-1}  % for review and submission
\usepackage{graphicx}  % needed for figures
\usepackage{dcolumn}   % needed for some tables
\usepackage{bm}        % for math
\usepackage{amssymb}   % for math
\usepackage{atlasphysics}
\usepackage{hyperref} 
\def\integLumiElec{4.9} % fb^-1
\def\integLumiMuon{5.0}

\def\expLimitMinusElec{13.8} % TeV
\def\expLimitMinusMuon{12.7}
\def\expLimitMinusComb{15.0}

\def\expLimitPlusElec{10.4}
\def\expLimitPlusMuon{9.9}
\def\expLimitPlusComb{11.3}

\def\obsLimitMinusElec{12.1}
\def\obsLimitMinusMuon{12.9}
\def\obsLimitMinusComb{13.9}

\def\obsLimitPlusElec{9.5}
\def\obsLimitPlusMuon{9.6}
\def\obsLimitPlusComb{10.2}

\def\bbbar{\ensuremath{b\bar{b}}}
\def\ccbar{\ensuremath{c\bar{c}}}
\def\ttbar{\ensuremath{t\bar{t}}}
\def\Zgstr{\ensuremath{Z/\gamma^{\ast}}}

\usepackage{preprintcover}  % load package
\PreprintCoverPaperTitle{Search for contact interactions and large extra dimensions in dilepton events from
 $pp$ collisions at $\sqrt{s} = 7$~\TeV{} with the ATLAS detector}  % paper title
\PreprintIdNumber{CERN-PH-EP-2012-279}  % CERN preprint number
\PreprintCoverAbstract{A search for nonresonant new phenomena, 
originating from either contact interactions or large extra spatial dimensions, 
has been carried out using events with two isolated electrons or muons. 
These events, produced at the LHC in proton-proton collisions at $\sqrt{s} = 7$~\TeV{}, were recorded by the ATLAS detector.
The data sample, collected throughout 2011, corresponds to an integrated luminosity of
$\integLumiElec$ and $\integLumiMuon$ fb$^{-1}$ in the $e^+ e^-$ and $\mu^+ \mu^-$
channels, respectively.
No significant deviations from the Standard Model expectation are observed.
Using a Bayesian approach, 95\% confidence level lower limits 
ranging from 9.0 to \obsLimitMinusComb~\TeV{}
are placed on the energy scale of
$\ell\ell q q$ contact interactions in the left-left isoscalar model.
Lower limits ranging from 2.4 to 3.9 \TeV{}
are also set on the string scale in large extra dimension models.
After combining these limits with results from a similar search in the diphoton channel, 
slightly more stringent limits are obtained.
}  % paper abstract
\PreprintJournalName{Phys. Rev. D}  % journal name
\begin{document}

\widetext

\title{Search for contact interactions and large extra dimensions in dilepton events from
 $pp$ collisions at $\sqrt{s} = 7$~\TeV{} with the ATLAS detector}
\author{The ATLAS Collaboration}
%\date{\today}

\begin{abstract}
A search for nonresonant new phenomena, 
originating from either contact interactions or large extra spatial dimensions, 
has been carried out using events with two isolated electrons or muons. 
These events, produced at the LHC in proton-proton collisions at $\sqrt{s} = 7$~\TeV{}, were recorded by the ATLAS detector.
The data sample, collected throughout 2011, corresponds to an integrated luminosity of
$\integLumiElec$ and $\integLumiMuon$ fb$^{-1}$ in the $e^+ e^-$ and $\mu^+ \mu^-$
channels, respectively.
No significant deviations from the Standard Model expectation are observed.
Using a Bayesian approach, 95\% confidence level lower limits 
ranging from 9.0 to \obsLimitMinusComb~\TeV{}
are placed on the energy scale of
$\ell\ell q q$ contact interactions in the left-left isoscalar model.
Lower limits ranging from 2.4 to 3.9 \TeV{}
are also set on the string scale in large extra dimension models.
After combination of these limits with results from a similar search in the diphoton channel, 
slightly more stringent limits are obtained.
\end{abstract}

\maketitle

%\linenumbers

\section{\label{sec:intro}Introduction}

Extensions to the Standard Model (SM), such as quark/lepton compositeness and large extra dimensions,
predict modifications to the SM dilepton invariant mass spectra. 
This paper presents a comparison of the number of expected and observed events at
high mass in the dielectron and dimuon datasets collected by the ATLAS 
detector~\cite{Aad:2008zzm} in 2011.
These events resulted from proton-proton collisions 
produced at $\sqrt{s} = 7$ \TeV{} by the LHC~\cite{Evans:2008zzb}.
The data are interpreted in the context of contact interactions (CI) and 
virtual graviton exchange in the Arkani-Hamed--Dimopoulos--Dvali (ADD) model~\cite{ArkaniHamed:1998rs}.

In the SM, quarks and leptons are fundamental particles.
However, if they are composite particles, with at least one common constituent, 
the interactions of these constituents would manifest themselves 
through an effective four-fermion contact interaction at energies
well below the compositeness scale. 
This type of contact interaction could also describe a new force with a messenger 
too heavy for direct observation at the LHC,
in analogy with Fermi's nuclear $\beta$~decay theory~\cite{Fermi:1934hr}.

The Lagrangian for a general contact interaction has the form~\cite{Eichten:1983hw}:

\begin{equation}\label{lagrangian}
\begin{array}{r@{\,}c@{}c@{\,}l@{\,}l}
\mathcal L = \frac{g^2}{2\Lambda^2}\;[&& \eta_{\rm LL}&\, \overline\psi_{\rm L}\gamma_{\mu}\psi_{\rm L}\,\overline\psi_{\rm L}\gamma^{\mu}\psi_{\rm L} \nonumber \\
& +&\eta_{\rm RR}& \overline\psi_{\rm R}\gamma_{\mu}\psi_{\rm R} \,\overline\psi_{\rm R}\gamma^{\mu}\psi_{\rm R} \\
&+&2\eta_{\rm LR}& \overline\psi_{\rm L}\gamma_{\mu}\psi_{\rm L} \,\overline\psi_{\rm R}\gamma^{\mu}\psi_{\rm R}&]\: ,\nonumber
\end{array}
\end{equation}
where $g$ is a coupling constant chosen so that $g^2/4\pi = 1$;
$\Lambda$ is the contact interaction scale, which in the context of compositeness
models, is the energy scale below which fermion constituents are bound;
and $\psi_{\rm L,R}$ are left-handed and right-handed fermion fields, respectively.
The parameters $\eta_{ij}$, where $i$ and $j$ are L or R (left or right), 
define the chiral structure of the new interaction.
Specific models are constructed by assigning particular combinations of these parameters
to be $-1$, $0$ or $+1$. For example, the left-left isoscalar model (LLIM) is defined 
by setting $\eta_{\rm LL} = \pm 1$ and $\eta_{\rm RR}$ = $\eta_{\rm LR}$ = 0.
The LLIM model, commonly used as a benchmark for contact interaction searches~\cite{Beringer:1900zz}, 
is utilized in this analysis.

The addition of the contact interaction Lagrangian to that of the SM  
modifies the Drell--Yan (DY) production cross section 
($q \bar{q} \to Z/\gamma^\ast \to \ell^+\ell^-$).
The largest deviations in the dilepton invariant mass spectra, either constructive 
or destructive, are expected at high mass and are determined by the sign of the 
parameter $\eta_{ij}$ and the scale $\Lambda$.
The differential cross section for the process $q \bar{q} \to \ell^+\ell^-$, 
including a contact interaction, can be separated into three components: 
a SM DY term, a pure contact interaction term ($F_{\rm C}$)
and a DY-CI interference ($F_{\rm I}$) term:
\begin{equation}
\frac{d\sigma}{dm_{\ell\ell}} = \frac{d\sigma_{\rm DY}}{dm_{\ell\ell}}
 - \eta_{\rm LL}\frac{F_{\rm I}(m_{\ell\ell})}{\Lambda^2} + \frac{F_{\rm C}(m_{\ell\ell})}{\Lambda^4}~,
\label{cross_section_CI}
\end{equation}
where $m_{\ell\ell}$ represents the final-state dilepton mass.
The full form of this expression is given in Ref.~\cite{Eichten:1984eu}.
Constructive (destructive) interference corresponds to
$\eta_{\rm LL} = -1$ ($+1$).
At the largest $\Lambda$ values to which this analysis is sensitive,
both interference and pure contact interaction terms play significant roles.
For example, at dilepton masses greater than 400~\GeV{} and $\Lambda = 12$~\TeV{},
the magnitude of the middle term in Eq.~(\ref{cross_section_CI}), which depends
on the interference, is about twice that of the last term.

Nonresonant deviations in the high mass dilepton invariant mass spectra 
are also predicted in large extra dimension models. 
These models were introduced to address some of the major
unresolved issues in particle physics such as the hierarchy problem.
The latter deals with the question of why gravity appears weak in comparison to the other three
SM interactions and why the electroweak scale ($\sim$1~\TeV) is 16 orders of magnitude smaller than 
the Planck scale ($M_{\rm Pl} \simeq 10^{16}$~\TeV).
Arkani-Hamed, Dimopoulos, and Dvali addressed these issues by postulating the existence of $n$ flat additional
 spatial dimensions of common size $R$, compactified on an $n$-dimensional torus~\cite{ArkaniHamed:1998rs}. 
The fundamental Planck scale in (4+$n$)-dimensional spacetime, $M_{\rm D}$, is then related 
to the scale $M_{\rm Pl}$ by Gauss's law: $M^{2}_{\rm Pl}$ = $M_{\rm D}^{n+2}$$R^{n}$. 
Consequently, the hierarchy problem can be solved with a Planck scale resulting from a 
fundamental scale ($M_{\rm D}^{n+2}$) near 1 \TeV{}
if the volume ($\propto R^{n}$) is large enough.

In the ADD model, the SM particles and their interactions 
are confined to a three-dimensional 
slice of the multidimensional world, but gravity permeates the additional 
dimensions of size $R$. This results in Kaluza-Klein (KK) modes of the graviton. 
The mass splitting of these KK modes is determined by the factor $1/R$. 
Resolution of the hierarchy problem
necessitates large extra dimensional volumes and consequently implies small values of $1/R$. 
This results in an almost continuous spectrum of KK graviton states
and hence a nonresonant increase in the expected rate of dilepton 
events at large invariant mass.
Performing the sum over the KK modes in the virtual graviton exchange process leads to an integral
which has to be regulated by an ultraviolet cutoff value ($\Lambda_{\rm UV}$).
The ADD model is a low-energy effective theory valid below the scale of the onset of 
quantum gravity, characterized by the scale $M_{\rm S}$.
The convention used throughout this analysis is to equate 
the cutoff to the scale of the effective theory ($\Lambda_{\rm UV}$=$M_{\rm S}$). 

For virtual graviton exchange, 
it is standard practice to present limits on the size of the 
extra dimensions in terms of $M_{\rm S}$, taken to be the string scale,
which is related to $M_{\rm D}$ by the following expression~\cite{Gleisberg:2003ue}:
\begin{equation}
M_{\rm S} = 2 \sqrt{\pi} \left[ \Gamma\left( \frac{n}{2} \right) \right]^{1/(n+2)} M_{\rm D}~.
\label{EqMSMD}
\end{equation}

The strength of gravity in the presence of extra dimensions is typically parametrized 
by $\eta_{\rm G}$ = $\mathcal{F}$/$M^{4}_{\rm S}$, where $\mathcal{F}$ is a dimensionless parameter 
of order unity.
The definition of $\mathcal{F}$ depends on the formalism chosen~\cite{Gleisberg:2003ue},
with three popular conventions: 
Giudice-Rattazzi-Wells (GRW)~\cite{Giudice:1998ck}, Hewett~\cite{Hewett:1998sn} 
and Han-Lykken-Zhang (HLZ)~\cite{Han:1998sg}. The different values are

\begin{eqnarray}\label{ADDformalism}
\mathcal F & = & 1,  {\rm ~~(GRW)}  \nonumber \\
\mathcal F & = & \frac{2 \lambda}{\pi} = \frac{\pm 2}{\pi}, {\rm ~~(Hewett)} \\
\mathcal F & = & \frac{2}{n-2} {\rm ~for~} n > 2. {\rm ~~(HLZ)} \nonumber
\end{eqnarray}
In the GRW and HLZ representations,
gravitational effects interfere constructively with the SM processes, while in 
Hewett's convention there can be destructive or constructive interference. 
This is encapsulated in the parameter $\lambda$, which is equal to $+1$ ($-1$) for 
constructive (destructive) interference.

The total cross section ($\sigma_{\rm tot}$), including effects of $q \overline{q}$-
and $g g$-initiated virtual graviton exchange, may be parametrized as
\begin{eqnarray}\label{cross_section_ADD}
\sigma_{\rm tot} = \sigma_{\rm SM} + \eta_{\rm G}F_{\rm int} + \eta_{\rm G}^{2}F_{\rm G},
\end{eqnarray}
where $\sigma_{\rm SM}$ is the SM cross section for the process being considered, and
$F_{\rm int}$ and $F_{\rm G}$ are functions of the cross sections involving the interference 
and pure graviton effects, respectively.
Note that the interference term 
has a linear dependence on $\eta_{\rm G}$, and therefore a quartic dependence on 
$M_{\rm S}$ (i.e., $\eta_{\rm G} \propto 1/M_{\rm S}^{4}$), whereas the pure graviton exchange term is quadratic 
in $\eta_{\rm G}$ and therefore has a $1/M_{\rm S}^{8}$ dependence. 
A study of signal yields in the kinematic range relevant to this analysis 
shows that the pure graviton term dominates those yields. 
This is in part due to the fact that the $gg$-initiated contribution to the graviton exchange 
process does not interfere with the $q \overline{q}$-initiated DY process.
Results are nevertheless presented for both $1/M_{\rm S}^4$ and $1/M_{\rm S}^8$ priors.

  Previous searches for contact interactions have been carried out 
in neutrino-nucleus and electron-electron scattering~\cite{McFarland:1997wx,Anthony:2005pm},
as well as at 
electron-positron~\cite{Schael:2006wu,Abdallah:2005ph,Abdallah:2008ab,Acciarri:2000uh,Abbiendi:2003dh}, 
electron-proton~\cite{Aaron:2011mv,Chekanov:2003pw}, and 
hadron colliders~\cite{Abe:1997gt,Affolder:2001gr,Abulencia:2006iv,Abbott:1998rr,Abazov:2009ac,Khachatryan:2011as,Aad:2011aj,Aad:2011tq}.
In the case of $e e q q$ contact interactions, the limits in the LLIM
for all quark flavors from $e^+ e^-$ experiments are
$\Lambda^- > 7.2$~\TeV{} and
$\Lambda^+ > 12.9$~\TeV{}~\cite{Schael:2006wu} at 95\% confidence level (C.L.) for
$\eta_{\rm LL} = -1$ and $+1$, respectively.
These limits assume that contact interactions of electrons with all quark
flavors are of the same strength.
The best limits set in the specific case of first generation quarks are
$\Lambda^- > 9.1$~\TeV{} and
$\Lambda^+ > 8.6$~\TeV{}~\cite{Abbiendi:2003dh} at 95\% C.L.
In the case of $ee q q$ contact interactions, the best limit for
constructive interference is $\Lambda^- > 10.1$~\TeV{}
from the ATLAS analysis of the first 1~\ifb\ of 2011 data~\cite{Aad:2011tq}.
The best limits in the case of $\mu\mu qq$ contact interactions are from an
analysis of the same data:
$\Lambda^- > 8.0$~\TeV{} and $\Lambda^+ > 7.0$~\TeV{}~\cite{Aad:2011tq}.

Previous searches for large extra dimensions in the ADD model via virtual graviton exchange
have been performed at 
electron-positron~\cite{Acciarri:1999bz,Acciarri:1999jy,Abreu:2000ap,Abreu:2000kh,Abbiendi:2002je,Abbiendi:1999wm},
electron-proton~\cite{Adloff:2000dp, Chekanov:2003pw},
and hadron colliders~\cite{Abazov:2008as,Abazov:2005tk,Abazov:2009ac,Chatrchyan:2012kc,Chatrchyan:2011jx,Chatrchyan:2011fq,ATLAS:2011ab,Aad:2012cy}.
Presently, the most stringent mass limits in the
dielectron and dimuon channels require $M_{\rm S} >$ 2.8~\TeV{} for each channel 
and 3.1~\TeV{} when combined
(in the GRW formalism with no $K$ factor)~\cite{Chatrchyan:2012kc}. 
The best limits to date arise
from the combination of these dilepton results with those from a search in the diphoton
final state, which increases the limit by $\sim$ 0.1~\TeV{}~\cite{Chatrchyan:2012kc}.
The following sections describe the first virtual graviton exchange search performed by ATLAS using dilepton data and its combination with an ATLAS diphoton 
data search~\cite{Aad:2012cy}.

\section{\label{sec:atlas}ATLAS Detector}
ATLAS is a multipurpose particle detector
composed of three main subsystems: the inner tracking detector, the calorimeter system 
and the muon spectrometer.
The inner detector is used to track charged particles within 
a pseudorapidity $\eta$\protect\footnote{ATLAS uses a right-handed coordinate system 
with its origin at the nominal
interaction point (IP) in the center of the detector and the $z$-axis along
the beam pipe. The $x$-axis points from the IP to the center of the LHC ring,
and the $y$-axis points upward. Cylindrical coordinates $(r,\phi)$
are used in the transverse plane, $\phi$ being the azimuthal angle around
the beam pipe. The pseudorapidity is defined in terms of the polar
angle $\theta$ as $\eta$~=~$-$~$\ln\tan(\theta /2)$.}
in the range $|\eta|$ $<$ 2.5.
It comprises a silicon pixel detector, 
a silicon-strip tracker and a transition radiation tracker. 
An axial 2~T magnetic field is produced by a superconducting solenoid.

The calorimeter system, covering $|\eta|$ $<$ 4.9, surrounds the solenoid and provides three-dimensional reconstruction of electromagnetic and hadronic showers. 
The lead/liquid-argon electromagnetic sampling calorimeter covers $|\eta| < 2.5$ 
and is finely segmented with a readout granularity varying by layer and 
with cells as small as $0.025 \times 0.025$ in ($\eta$, $\phi$) 
to provide precise energy and position resolution, as needed for
electron and photon identification and energy measurement.
Hadron calorimetry is provided by an 
iron/scintillator tile calorimeter in the central pseudorapidity range  $|\eta| < 1.7$ 
and a lead/liquid-argon calorimeter extending the pseudorapidity range up to $|\eta| = 3.2$.
Both the electromagnetic and hadronic calorimeters have liquid-argon-based forward detectors, with copper or tungsten
as an absorber, to extend coverage up to $|\eta| = 4.9$.

Outermost is the muon spectrometer, another key detector component for this analysis. 
Three layers of precision tracking chambers, comprising monitored drift tubes 
and cathode strip chambers, enable muon reconstruction up to $|\eta| = 2.7$. 
The magnetic field is provided by three large air-core toroidal magnet systems 
(one barrel and two end caps), each consisting of eight
azimuthally symmetric superconducting coils. 
Triggering capability up to $|\eta| = 2.4$ is provided by fast resistive plate 
chambers in the barrel and thin-gap chambers in the end caps.

\section{\label{sec:sim}Signal and Background Modeling}
The dominant background contribution comes from the SM DY process with
smaller contributions from $\ttbar$ and
electroweak diboson ($WW$, $WZ$, and $ZZ$) production.
In the dielectron channel, there is also a significant background
from multijet and $W$+jets events in which jets are misidentified as electrons.
Backgrounds are estimated using fully simulated Monte Carlo (MC) samples
except for the combined multijet and $W$+jets background, which is determined
from the data.

DY samples are generated with {\sc Pythia} 6.421~\cite{Sjostrand:2006za} using
{\sc MRST2007} LO$^{\ast\ast}$ parton distribution functions (PDFs)~\cite{Sherstnev:2007nd,*Sherstnev:2008dm}.
The diboson background is generated with {\sc Herwig} 6.510~\cite{Corcella:2000bw} using 
{\sc MRST2007} LO$^{\ast\ast}$ PDFs.
For the $\ttbar$ background, event generation is performed with 
{\sc Mc@nlo} 4.01~\cite{Frixione:2002ik} 
and the CTEQ~6.6 PDFs~\cite{Nadolsky:2008zw},
as well as {\sc Herwig} to model the underlying event and parton showers.
Production of diboson and $\ttbar$ events relies on 
{\sc Jimmy} 4.31~\cite{Butterworth:1996zw} to describe multiple interactions.

For the contact interactions analysis, {\sc Pythia} 6.421 and the
{\sc MRST2007} LO$^{\ast\ast}$ PDFs are used to generate
both DY and CI as part of the same process in order to properly handle
the interference between DY and CI contributions, as shown in Eq.~(\ref{cross_section_CI}).
All quark flavors contribute to the DY+CI production.

Final-state radiation is simulated with {\sc Photos}~\cite{Golonka:2005pn} for all
the samples listed above.
Higher-order QCD corrections are implemented via mass-dependent $K$ factors
defined as the ratio of the next-to-next-to-leading-order (NNLO) \Zgstr\ cross section
calculated using {\sc Phozpr}~\cite{Hamberg:1990np} and the {\sc MSTW2008} 
NNLO PDFs~\cite{Martin:2009iq},
to the LO \Zgstr\ cross section from {\sc Pythia}.
Higher-order electroweak corrections originating from virtual gauge-boson loops
are computed using the {\sc Horace} NLO event generator~\cite{CarloniCalame:2007cd}.
The mass-dependent QCD and electroweak $K$ factors are applied to both DY and DY+CI samples.

For the large extra dimensions analysis, 
{\sc Sherpa} 1.3.1~\cite{Gleisberg:2008ta} and the {\sc CTEQ6L} PDFs~\cite{Pumplin:2002vw}
are used to generate DY+ADD events at leading order. 
The generated dilepton mass is required to be less than the scale $M_{\rm S}$,
which is chosen to be in the range between 2 and 3~\TeV{} in this study, 
since the model is not valid at energies beyond that scale.

Finally, the generated samples are processed through a full simulation of the
ATLAS detector~\cite{Aad:2010ah} based on {\sc Geant}~4~\cite{Agostinelli:2002hh,*Allison:2006ve}
and reconstructed with the same software as for the data.
Several corrections derived from data control samples are applied to the simulated samples. 
Specifically, the energy scale and resolution for electrons 
are corrected so that the $Z \to ee$ mass distribution in simulation matches the data~\cite{Aad:2011mk}.
Similarly, the muon transverse momentum scale and resolution are adjusted to reproduce the muon tracking
performance measured in $Z \to \mu\mu$ data as well as several dedicated data samples~\cite{MuonPerf}.
The effect of pileup (multiple $pp$ interactions in the same or nearby bunch crossings)
is included by superimposing minimum bias events, in the same proportion as in data,
on top of the hard scattering process generated. 
Small corrections are included in the analysis to properly describe
the pileup conditions for the selected data sample.

\section{\label{sec:event}Event Selection and Comparison Between Expected and Observed Yields}

This analysis follows the same event selection as the search for 
new heavy resonances~\cite{Aad:2012hf} in the dielectron channel
but uses a somewhat tighter selection in the dimuon channel.

The data sample was collected during LHC operation in 2011 
and corresponds to a total integrated luminosity
of $\integLumiElec$ and $\integLumiMuon$ fb$^{-1}$
in the $e^{+}e^{-}$ and $\mu^{+}\mu^{-}$ final states, respectively. 
The events recorded by the ATLAS detector were selected by requiring that 
they pass specific triggers. The trigger for the dielectron dataset 
required the presence
of two electromagnetic clusters consistent with originating from electrons with
transverse momentum $p_{\rm T}$ above 20 \GeV{}, whereas events in the muon dataset were required to 
pass at least one of two single-muon triggers with $p_{\rm T}$ thresholds of 22 \GeV{} and
40 \GeV{}.

After passing the trigger selection, events are
required to have a pair of either electrons or muons
with $p_{\rm T}$ greater than 25 \GeV{}. 
Furthermore, events are required to be recorded during stable beam conditions and
with detector components operational.
To reject cosmic ray events and beam halo background, events are required to have a 
reconstructed vertex with at least three charged particle tracks with 
$p_{\rm T} > 0.4$ \GeV{}.
If more than one such vertex is found, the vertex with the largest $\Sigma p_{\rm T}^2$
is selected as the primary vertex of the event, where the sum is over all
charged particles associated with the given vertex.
Electron candidates are confined to $|\eta| < 2.47$, with
the calorimeter barrel--to--end-cap transition region 
1.37 $\leq$ $|\eta|$ $\leq$ 1.52 excluded
due to the degraded energy resolution in this region.
%Muon candidates are required to be within the inner detector acceptance $|\eta| < 2.5$.
No explicit $\eta$ requirement is placed on muon candidates, but 
the selection described below leads to negligible acceptance beyond $|\eta|$
of approximately 2.5.

Electron candidates are formed from clusters of cells in
the electromagnetic calorimeter where energy is deposited.
Identification criteria based on the transverse shower shape, 
the leakage into the hadronic calorimeter, and the association 
to an inner detector track are applied to the cluster to satisfy 
the {\it medium} electron definition~\cite{Aad:2011mk}.
The electron energy is obtained from the calorimeter measurements
and its direction from the associated inner detector track.
A hit in the first layer of the pixel detector is required (if an active
pixel module is traversed) to suppress background from photon conversions.
Further jet background suppression is achieved
by demanding that the highest-$p_{\rm T}$ electron in the event be isolated. 
To this effect, the sum of the transverse energies, $\Sigma E_{\rm T}$, 
in calorimeter cells
within a radius $R = \sqrt{(\Delta\eta)^2 + (\Delta\phi)^2}$ of 0.2 around the electron
direction, is required to be less than 7~\GeV{}.
The core of the electron energy deposition is excluded and the
sum is corrected for transverse shower leakage and pileup.
The two electron candidates are not required to have opposite 
charge because of possible charge misidentification either 
due to bremsstrahlung or limited momentum resolution of the inner 
detector at high $p_{\rm T}$.
If the event contains more than two selected electrons, the two
electrons with the highest-$p_{\rm T}$ sum are chosen.
For these selection criteria, the overall event acceptance for DY events
has a small dependence on the dielectron mass above 500 \GeV{}, with
a value of approximately 65\% at 1~\TeV{}.

Muon candidates are reconstructed independently in the inner detector and the muon spectrometer. The momentum is taken from a combined fit to the measurements from the two subsystems. 
To obtain optimal momentum resolution and accurate modeling by the simulation, 
muon candidates are required to have at least three hits in each of the inner, 
middle, and outer detector layers of the muon spectrometer,
and to have at least one hit in each of two different layers in the nonbending $xy$ plane.
To suppress background from cosmic rays, requirements are imposed on the
primary vertex (PV) position and the muon impact parameter relative to the PV: 
$z$ coordinate of the PV $|z_{\rm PV}| <$ 200~mm,
muon transverse impact parameter $|d_0| <$ 0.2~mm and
muon $z$ coordinate $|z_0 - z_{\rm PV}|<$ 1~mm.
Furthermore, the muons are required to be isolated to reduce background from jets: 
$\Sigma p_{\rm T} (R < 0.3) / p_{\rm T}(\mu) <$ 0.05,
where the sum is over inner detector tracks within a radius of 0.3 around the muon direction.
If more than one opposite-sign muon pair is found in an event,
the pair with the highest-$p_{\rm T}$ sum is chosen. 
The overall event acceptance for DY events
has only a weak dependence on the dimuon mass, with
a value of approximately 40\% at 1~\TeV{}.
This is lower than the acceptance in the dielectron channel primarily due to 
the stringent requirements on the presence of hits in all three layers of 
the muon spectrometer and the extent of the three-layer geometrical coverage.

The $W$+jets background in the dimuon channel is estimated from 
simulated samples and is found to be negligible since the event must
contain two well-measured high-$p_{\rm T}$ isolated muons.
Likewise, the multijet background, estimated directly from the data
by reversing the muon isolation criterion, is found to be negligible.
The multijet and $W$+jets backgrounds are not negligible in the dielectron channel.
They are estimated primarily from the data using several methods~\cite{Aad:2012hf}.
The first method determines the multijet background from the data and
relies on the MC simulation for the $W$+jets contribution. The background is measured
with a template built by reversing one of the electron identification criteria
and normalized to data in the range $70 < m_{ee} < 200$ \GeV.
Another independent method that is sensitive to both multijet and $W$+jets backgrounds
uses jet-enriched data samples either from jet triggers or from 
the same trigger used to select the events in this analysis.
The method relies on jet misidentification rates, defined as the number of jets
that pass the full electron selection divided by the number that pass a loose electron selection obtained
by reversing one of the identification criteria.
The background estimate is then constrained by a fit in the range $140 < m_{ee} < 850$ \GeV{}.
The final combined multijet and $W$+jets background is obtained
with a simple average of the expected
event yields from the different methods.

Extensive comparisons between data and MC simulation were performed
at the level of single-lepton distributions
to confirm that the simulation reproduces the selected data,
especially at high momentum.
Figure~\ref{pt} shows good data-MC agreement in
the lepton transverse momentum distributions for events passing all
selection criteria.
\begin{figure}[!tb]
  \includegraphics[width=0.48\textwidth]{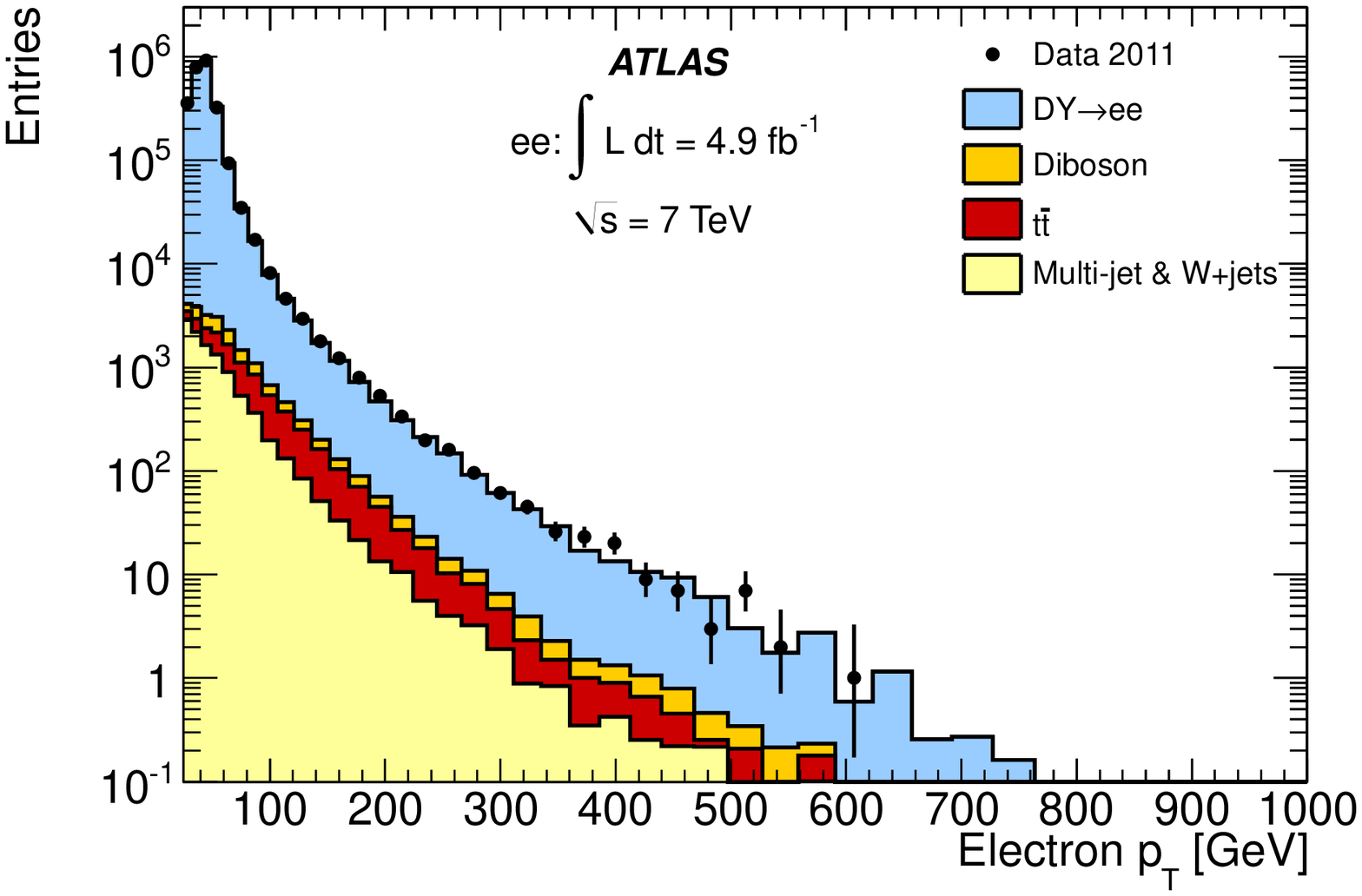}
  \includegraphics[width=0.48\textwidth]{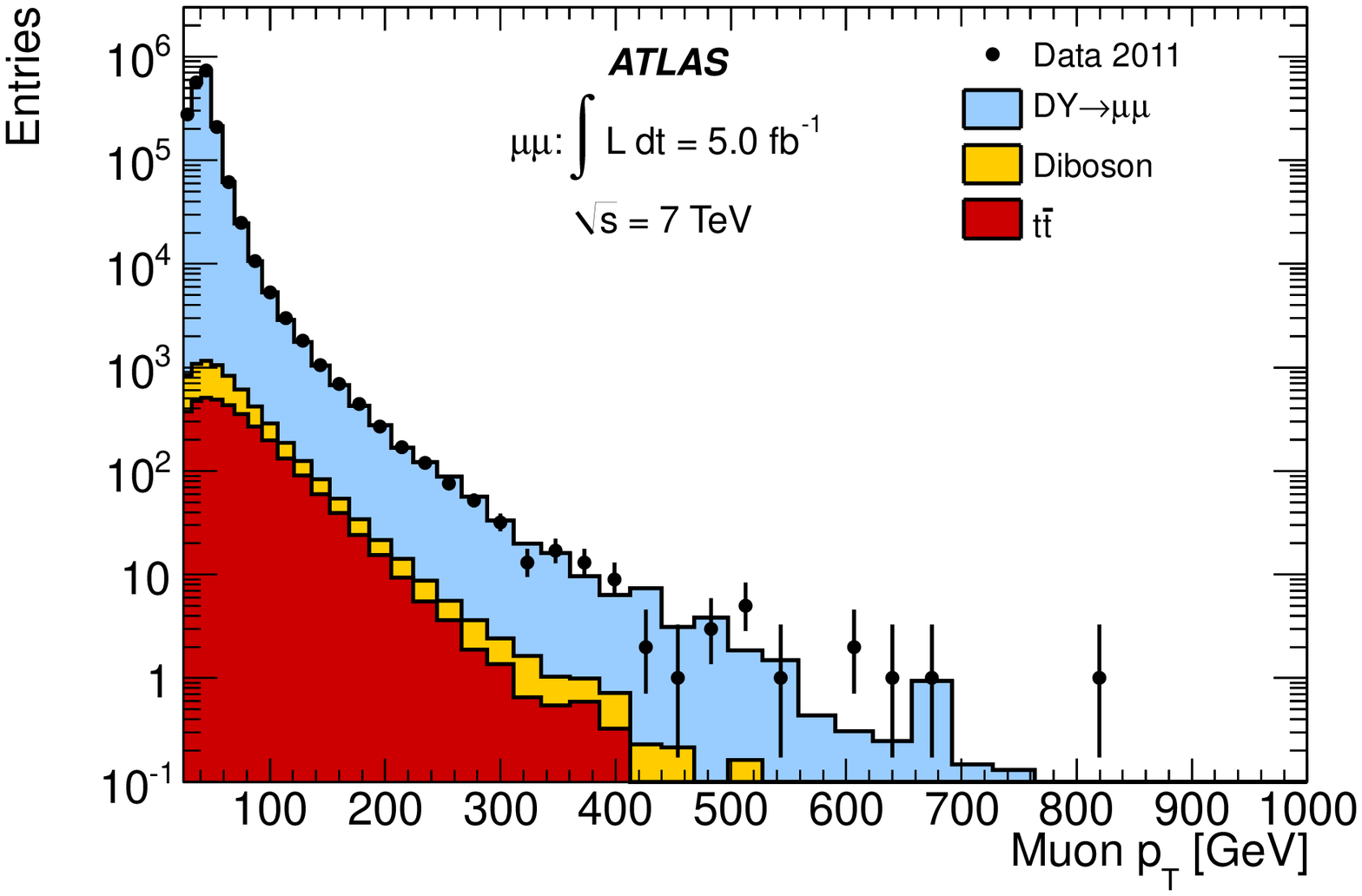}
  \caption{\label{pt} Lepton transverse momentum distributions in the dielectron (top panel)
  and dimuon (bottom panel) 
  channels for data (points) and Monte Carlo simulation (histograms).
  The bin width is constant in $\log(p_{\rm T})$.
  }
\end{figure}

Figure~\ref{mass} shows the dielectron and dimuon mass distributions for
selected events. Also shown are the predicted
contributions from SM and new phenomena (NP) for several choices of model parameters.
The expected SM distribution is dominated by the DY process over the entire
mass range and is found to describe the data well.
The level of agreement with the SM expectation is also illustrated
in Fig.~\ref{mass_integral}, which shows the number of events above a minimum
mass $m_{\ell\ell}^{\rm min}$.

\begin{figure}[!tb]
  \includegraphics[width=0.48\textwidth]{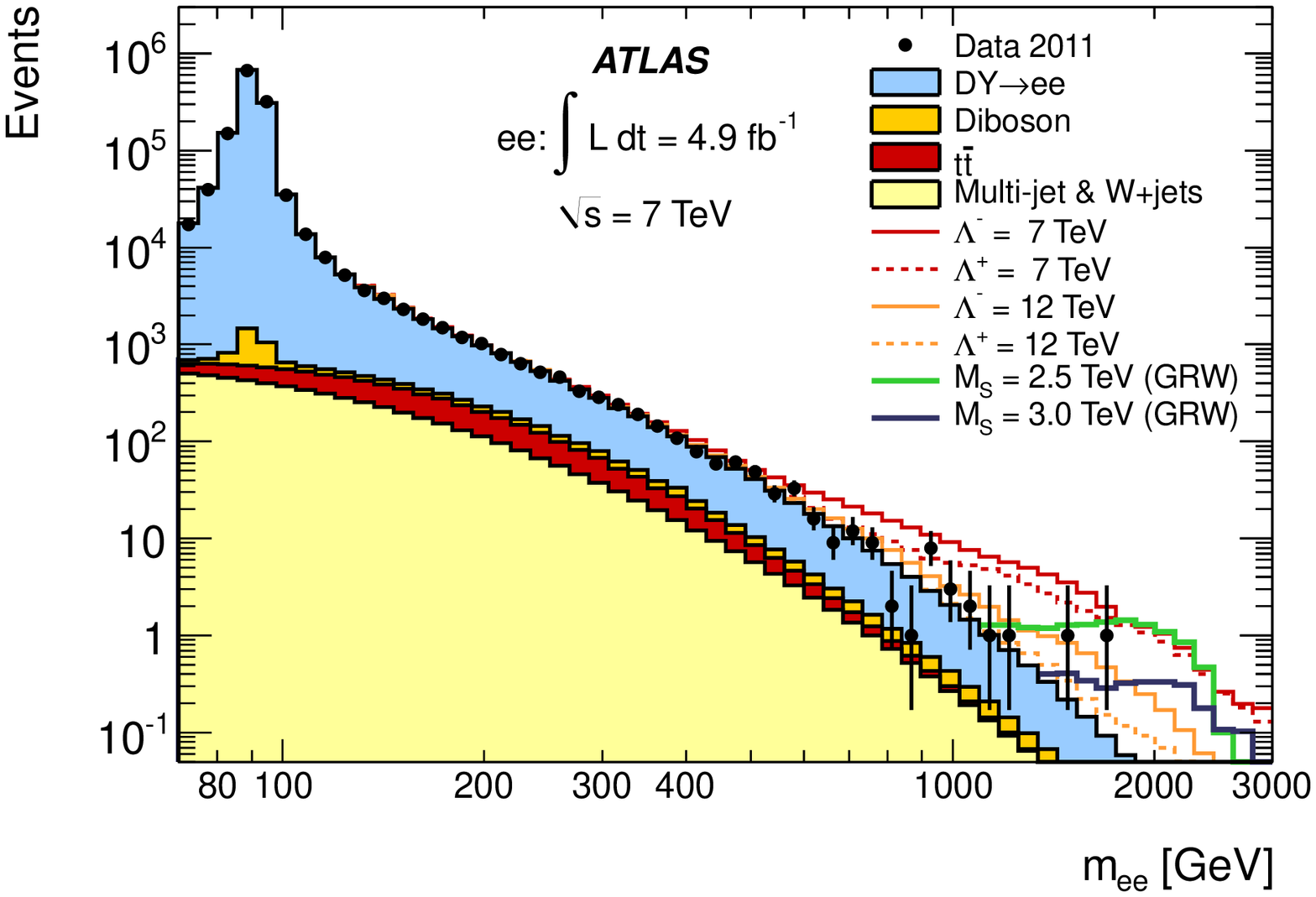}
  \includegraphics[width=0.48\textwidth]{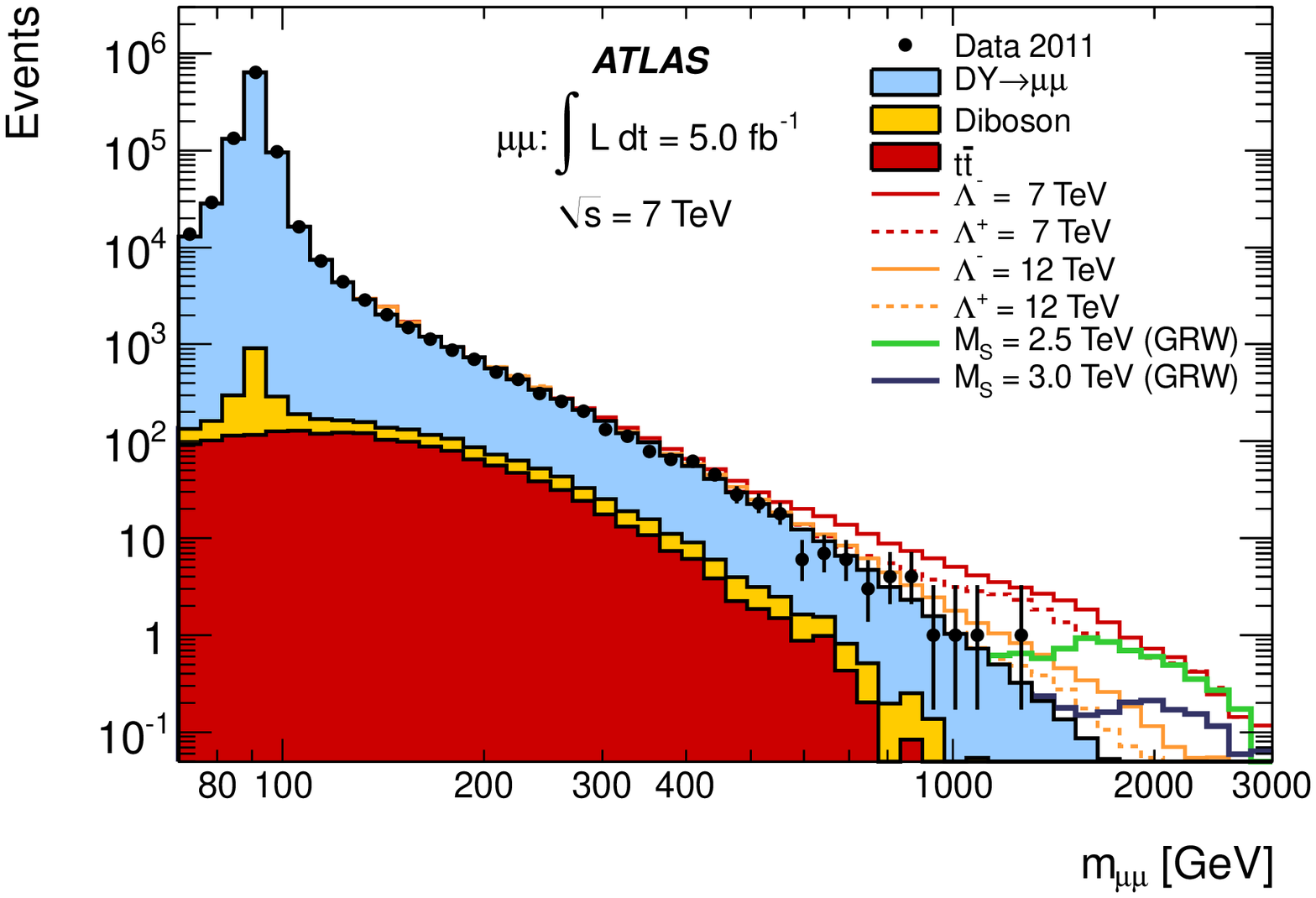}
  \caption{\label{mass} Dielectron (top panel) and dimuon (bottom panel) invariant mass distributions
  for data (points) and Monte Carlo simulation (filled histograms). The open histograms correspond to
  the distributions expected in the presence of contact interactions or large extra dimensions
  for several model parameters.
  The bin width is constant in $\log(m_{\ell\ell})$.
  }
\end{figure}
\begin{figure}[!tb]
  \includegraphics[width=0.48\textwidth]{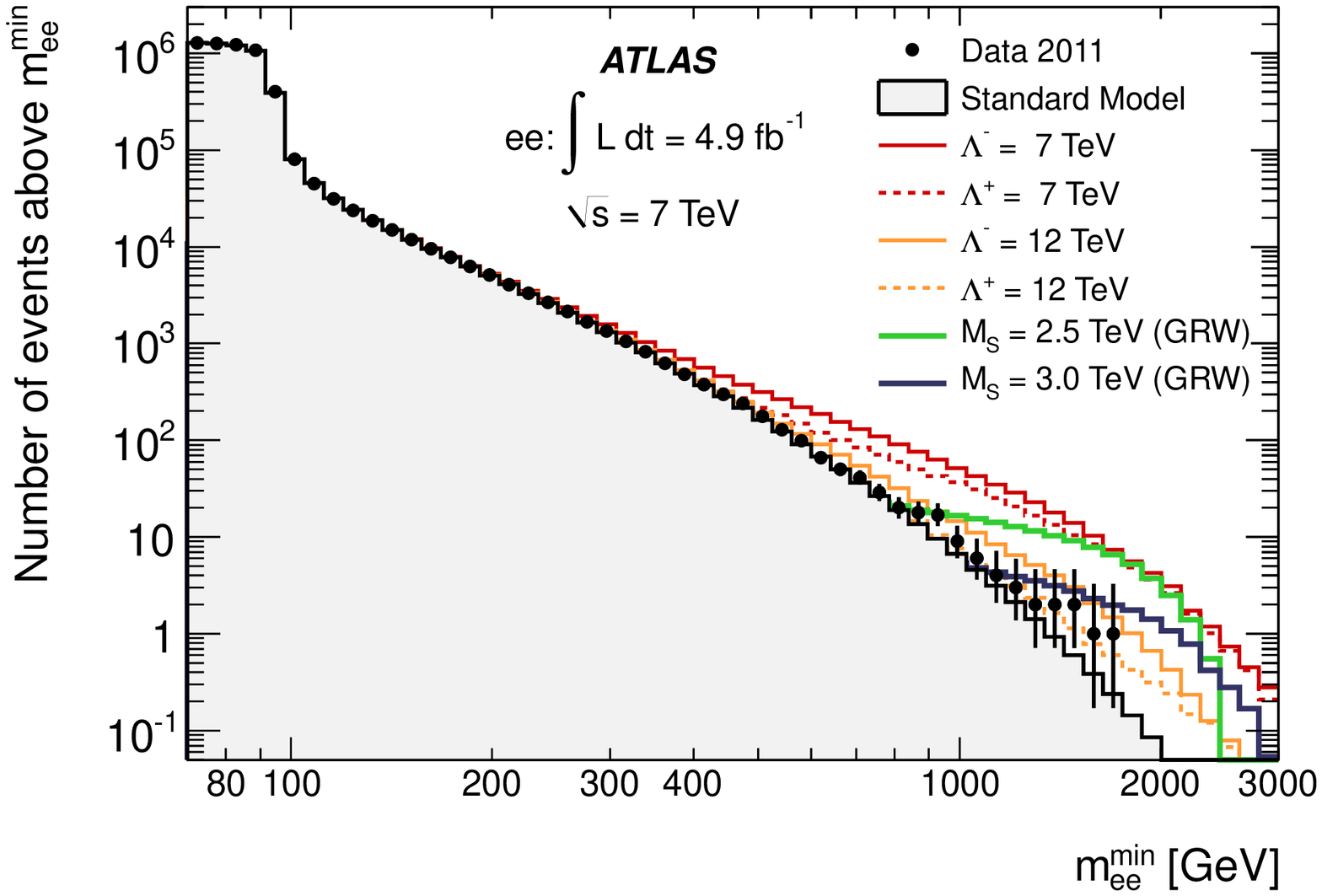}
  \includegraphics[width=0.48\textwidth]{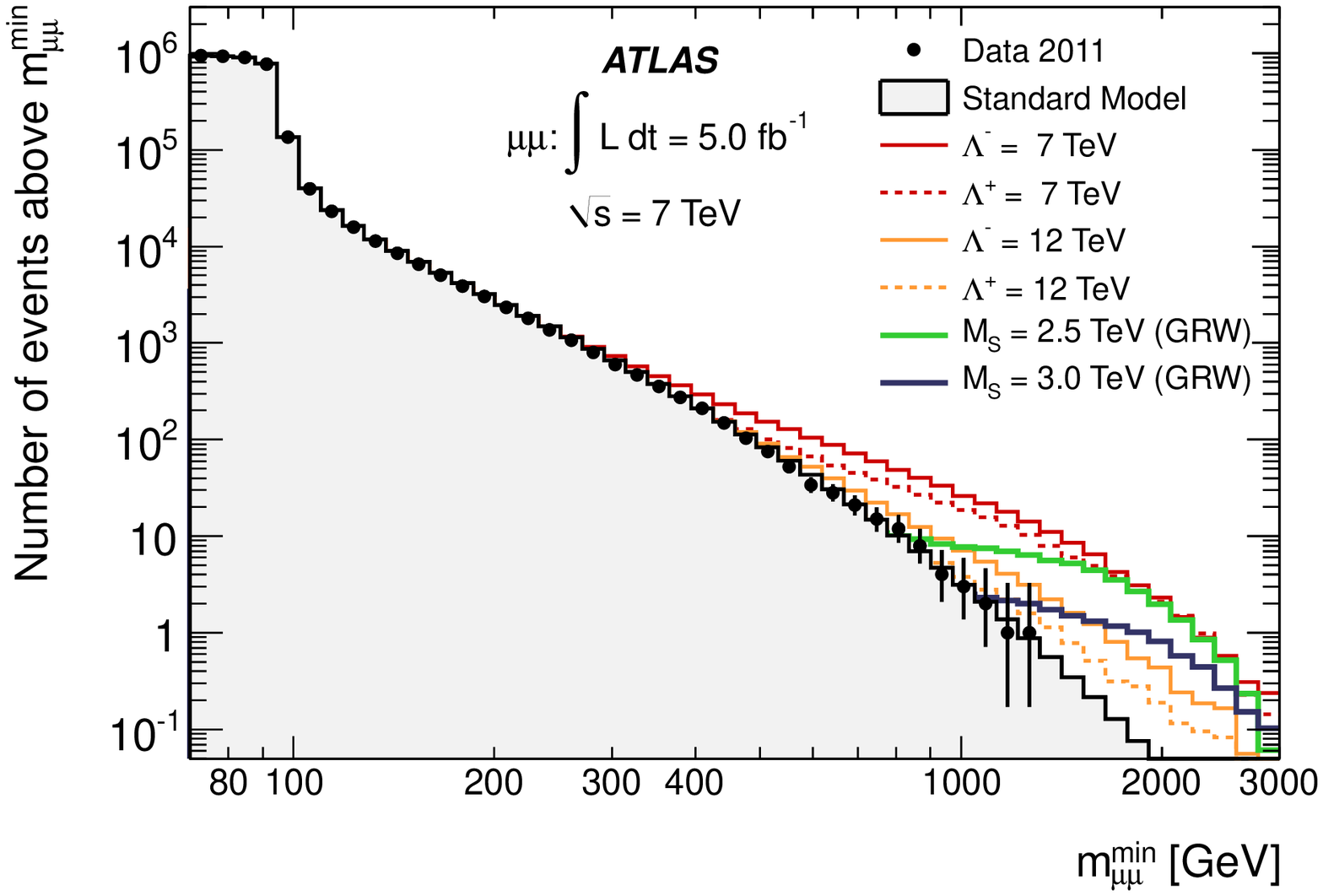}
  \caption{\label{mass_integral} Distribution of the number of events
  with dilepton mass above $m_{\ell\ell}^{\min}$ for data (points)
  and SM prediction from Monte Carlo simulation (filled histograms, shaded gray) in the dielectron channel (top panel)
  and dimuon channel (bottom panel). The open solid and dashed histograms correspond to the expected distributions
  in the presence of contact interactions or large extra dimensions for several
  model parameters.
  The bin width is constant in $\log(m_{\ell\ell}^{\min})$.
  }
\end{figure}
A more quantitative comparison is provided in Tables~\ref{yields_elec}
and \ref{yields_muon} showing the numbers of observed and expected events
in the dielectron and dimuon channels, respectively.
The expected yields are normalized to the number of events observed in the $Z$ peak
control region ($70 < m_{\ell\ell} < 110$ \GeV{}).
The mass region shown in these tables corresponds to the CI search region
defined by $m_{\ell\ell} > 400$ \GeV.
These tables also display the expected yields for the SM+CI signal for the two scenarios where the CI interferes either constructively or destructively with the SM.

\begin{table*}[!tb]
\newcommand\T{\rule{0pt}{2.6ex}}
\newcommand\B{\rule[-1.2ex]{0pt}{0pt}}
\caption{Expected and observed numbers of events in the dielectron channel for the
 contact interactions search region. The yields are normalized to the $Z$ peak control region
 and include predictions for SM backgrounds as well as
 for SM+CI with different CI scales for constructive ($\Lambda^-$)
 and destructive ($\Lambda^+$) interference. 
 The errors quoted originate from both systematic uncertainties and limited MC statistics.
}
\label{yields_elec}
\centering
%\scriptsize
\vspace{0.2cm}
\begin{tabular}{l r@{\ $\pm$\ }l r@{\ $\pm$\ }l r@{\ $\pm$\ }l r@{\ $\pm$\ }l r@{\ $\pm$\ }l}
\hline
\hline
$m_{ee}$ [\GeV{}] \T \B & \multicolumn{2}{c}{400--550} & \multicolumn{2}{c}{550--800} & \multicolumn{2}{c}{800--1200} & \multicolumn{2}{c}{1200--1800} & \multicolumn{2}{c}{$>$1800} \\
\hline
DY \T           & 203 & 10 & 62.5 & 3.4 & 12.1 & 0.9 & 1.38 & 0.17 & 0.085 & 0.025\\
\ttbar \T       & 22.6 & 2.1 & 4.05 & 0.34 & 0.308 & 0.026 & \multicolumn{2}{c}{$<0.05$} & \multicolumn{2}{c}{$<0.01$} \\
Diboson \T      & 12.1 & 0.7 & 4.08 & 0.21 & 0.88 & 0.05 & 0.111 & 0.006 & 0.0100 & 0.0006 \\
Multijet/$W$+jets \T & 38   & 23  &  11 & 8   & 2.0  & 1.8 & 0.24 & 0.28 & 0.022 & 0.029\\
\hline
 Total background \T \B    & 276 & 25   &  82 & 9   & 15.3 & 2.0 & 1.74 & 0.33 & 0.12 & 0.04 \\
\hline
$\Lambda^-$~=~~3~\TeV{} \T & 1460 & 70 & 1400 & 80 & 1090 & 60 & 525 & 35 & 148 & 13 \\
$\Lambda^-$~=~~4~\TeV{} \T & 680 & 40 & 519 & 27 &  360 & 21 &  171 & 12 & 44 & 4 \\
$\Lambda^-$~=~~5~\TeV{} \T & 463 & 30 & 281 & 15 & 162 & 9 & 77 & 5 & 19.8 & 1.9 \\
$\Lambda^-$~=~~7~\TeV{} \T & 332 & 27 & 145 & 10 & 59 & 4 & 22.0 & 1.6 & 4.8 & 0.5 \\
$\Lambda^-$~=~12~\TeV{} \T \B & 293 & 27 & 96 & 9 & 23.6 & 2.3 & 5.1 & 0.5 & 0.87 & 0.14 \\
\hline
$\Lambda^+$~=~~3~\TeV{} \T & 1080 & 50 & 1120 & 60 & 920 & 50 & 493 & 33 & 128 & 11 \\
$\Lambda^+$~=~~4~\TeV{} \T & 484 & 30 & 373 & 20 & 291 & 17 & 156 & 10 & 40 & 4 \\
$\Lambda^+$~=~~5~\TeV{} \T & 342 & 27 & 182 & 11 & 114 & 6 & 61 & 4 & 18.3 & 1.6 \\
$\Lambda^+$~=~~7~\TeV{} \T & 268 & 27 & 102 & 10 & 37.4 & 2.6 & 15.1 & 1.0 & 4.3 & 0.4 \\
$\Lambda^+$~=~12~\TeV{} \T \B &  260 & 27 &  82 & 9 & 15.1 & 2.2 & 2.5 & 0.4 & 0.41 & 0.08 \\
\hline
Data \T \B & \multicolumn{2}{c}{270}  &  \multicolumn{2}{c}{88}  &  \multicolumn{2}{c}{17}  &  \multicolumn{2}{c}{3}  &  \multicolumn{2}{c}{0} \\
\hline
\hline
\end{tabular}
%}
\end{table*}

\begin{table*}[!tb]
\newcommand\T{\rule{0pt}{2.6ex}}
\newcommand\B{\rule[-1.2ex]{0pt}{0pt}}
\caption{Expected and observed numbers of events in the dimuon channel for the
 contact interactions search region. The yields are normalized to the $Z$ peak control region
 and include predictions for SM backgrounds as well as
 for SM+CI with different CI scales for constructive ($\Lambda^-$)
 and destructive ($\Lambda^+$) interference. 
 The errors quoted originate from both systematic uncertainties and limited MC statistics.
}
\label{yields_muon}
\centering
%\scriptsize
\vspace{0.2cm}
\begin{tabular}{l r@{\ $\pm$\ }l r@{\ $\pm$\ }l r@{\ $\pm$\ }l r@{\ $\pm$\ }l r@{\ $\pm$\ }l}
\hline
\hline
\T $m_{\mu\mu}$ [\GeV{}] \T \B & \multicolumn{2}{c}{400--550} & \multicolumn{2}{c}{550--800} & \multicolumn{2}{c}{800--1200} & \multicolumn{2}{c}{1200--1800} & \multicolumn{2}{c}{$>$1800} \\
\hline
DY    \T    &  123  & 6    &  37.4 & 2.2   &  7.1  & 0.6  &  0.82 & 0.11  &  0.058 & 0.022 \\ 
\ttbar \T   &  13.4 & 1.4  &  3.1  & 0.5   &  0.04 & 0.12 &  \multicolumn{2}{c}{$<0.05$}  &  \multicolumn{2}{c}{$<0.01$} \\ 
Diboson \T  &  7.9  & 0.4  &  2.66 & 0.15  &  0.55 & 0.04 & 0.075 & 0.006 & 0.0124 & 0.0031 \\ 
\hline
Total background \T \B &  145 & 6   &  43.2 & 2.2    &  7.7 & 0.6   &  0.89 & 0.11 &  0.070 & 0.022 \\ 
\hline
$\Lambda^-$~=~~3~\TeV{} \T & 870 & 50 & 770 & 50 & 580 & 40 & 296 & 28 & 82 & 22 \\
$\Lambda^-$~=~~4~\TeV{} \T & 405 & 19 & 301 & 17 & 201 & 14 & 87 & 8 & 27 & 7 \\
$\Lambda^-$~=~~5~\TeV{} \T & 256 & 12 & 159 & 8 & 94 & 6 & 41 & 4 & 12.7 & 3.4 \\
$\Lambda^-$~=~~7~\TeV{} \T & 184 & 9 & 79 & 4 & 30.1 & 1.9 & 12.3 & 1.2 & 2.9 & 0.8 \\
$\Lambda^-$~=~12~\TeV{} \T \B & 157 & 9 & 50.6 & 3.1 & 12.3 & 0.9 & 2.81 & 0.31 & 0.53 & 0.15 \\
\hline
$\Lambda^+$~=~~3~\TeV{} \T & 628 & 31 & 650 & 40 & 500 & 40 & 248 & 23 & 75 & 20 \\
$\Lambda^+$~=~~4~\TeV{} \T & 271 & 12 & 203 & 11 & 159 & 11 & 85 & 8 & 22 & 6 \\
$\Lambda^+$~=~~5~\TeV{} \T & 182 & 9 & 98 & 5 & 64 & 4 & 31.4 & 2.9 & 11.5 & 3.0 \\
$\Lambda^+$~=~~7~\TeV{} \T & 141 & 8 & 50.8 & 3.1 & 19.7 & 1.2 & 8.4 & 0.8 & 2.5 & 0.7 \\
$\Lambda^+$~=~12~\TeV{} \T \B & 140 & 8 & 40.2 & 3.0 & 7.4 & 0.7 & 1.57 & 0.20 & 0.25 & 0.08 \\
\hline
Data \T \B  & \multicolumn{2}{c}{151}  &  \multicolumn{2}{c}{36}  &  \multicolumn{2}{c}{9}  &  \multicolumn{2}{c}{1}  &  \multicolumn{2}{c}{0} \\
\hline
\hline
\end{tabular}
%}
\end{table*}

\section{\label{sec:systematics}Systematic Uncertainties}

Except for the multijet and $W$+jets background contributions to the dielectron channel,
all signal and background event yield estimates are based on MC simulation.
Because these yields are normalized in the $Z$ peak control region,
only mass-dependent systematic uncertainties affect the event yield estimates
in the higher-mass signal region.
The only exception is a 5\% uncertainty applied to the signal yield to account for the
uncertainty in the \Zgstr\ cross section which affects the signal normalization.

Experimental uncertainties arise from lepton energy/momentum scale and resolution,
as well as trigger, reconstruction and identification efficiencies.
In the dielectron channel, 
the largest experimental uncertainty comes from the
combined multijet and $W$+jets background estimate.
It is determined from the envelope of the three separate
methods used, including the effect of varying the mass ranges in the background fits
and the uncertainties in the $\eta$ and $p_{\rm T}$ dependence of the jet 
misidentification rates.
Electron energy scale and resolution are determined from
data via $J/\psi \to ee$ and $Z \to ee$ mass distributions, as well as 
studies of electron $E/p$ in $W \to e \nu$ decays~\cite{Aad:2011mk}.
The uncertainty in the constant term that dominates the
resolution at high energy has negligible impact on this analysis.
A somewhat larger impact comes from the energy scale knowledge, resulting
in a systematic error of 1.2\% and 2.4\% for dielectron masses of 1
and 2 \TeV, respectively.
A slight efficiency drop of 1.0\% per \TeV{} is predicted by the simulation
due to the isolation requirement on the leading electron.
To account for this effect, an uncertainty of the same magnitude is
introduced.

In the dimuon channel, the largest contribution to the experimental
systematic error comes from the muon reconstruction efficiency and muon
resolution.
A slight drop in reconstruction efficiency is predicted by 
the simulation at high $p_{\rm T}$
due to the presence of additional hits in the muon spectrometer from muons
undergoing large energy loss in the detector.
An uncertainty of 3.0\% (6.0\%) at a dimuon mass of 1 (2) \TeV\ is assessed, 
corresponding to the magnitude of this effect.
The limited knowledge of the momentum scale determined from $Z \to \mu\mu$ data
has a negligible impact on the analysis. 
The momentum resolution in the simulation is adjusted based on $Z \to \mu\mu$
and $W \to \mu \nu$ data, as well as dedicated straight muon track data
collected with the toroids turned off and tracks passing through
overlapping sectors in the muon spectrometer. The latter provide
two independent momentum measurements for the same muon. The toroid-off and overlapping
sector tracks are key to determining the muon reconstruction performance at
high $p_{\rm T}$.
The uncertainty in the muon resolution, taken as equal in
magnitude to the correction applied to the simulation, results in a change
in the event yield of 1.2\% (12\%) for $m_{\mu\mu} = 1~(2)$~\TeV{}.

The largest error contribution due to theory arises from limited knowledge
of the PDFs, $\alpha_{\rm S}$, and QCD $K$ factors.
Scale uncertainties are computed by taking the maximum deviations
obtained by independently varying the renormalization
($\mu_{\rm R}$) and factorization ($\mu_{\rm F}$) scales by a factor of 2 but with the
constraint that the ratio $\mu_{\rm F} / \mu_{\rm R}$ does not change by more than a factor of 2.
The $\alpha_{\rm S}$ and PDF uncertainties are determined with the MSTW2008NNLO eigenvector
PDF sets and the different PDFs corresponding to variations of $\alpha_{\rm S}$.
The overall uncertainty is computed using 90\% C.L. ranges 
and includes the envelope of the uncertainty bands for the following different
PDF sets: MSTW2008, NNPDF2.1, CT10, and CT10W.
PDFs are the largest source of uncertainty, with the envelope of all
PDFs considered becoming the dominant contribution above a few hundred \GeV.
Uncertainties in the electroweak $K$ factor~\cite{CarloniCalame:2006zq} originate from the
calculation of real boson radiation, ${\cal O}(\alpha \alpha_{\rm S})$ corrections,
higher-order electroweak corrections, an assumed uncertainty of 10\% in the
contribution from photon-induced processes, and a difference in the electroweak
renormalization scheme definition used in {\sc Pythia} and in the calculation of the electroweak
corrections with {\sc Horace}. The latter source is the largest
contribution to the electroweak uncertainty.

The systematic uncertainties are summarized in Table~\ref{tab:systematics}.
Although not explicitly listed in this table, the uncertainty due to limited 
MC statistics is also taken into account in the limit setting.
For DY+CI MC samples, this uncertainty grows from about 4\% at low $m_{\ell\ell}$
to about 30\% at high $m_{\ell\ell}$ for $\Lambda = 12$ \TeV{}.

\begin{table*}[!tb]
\newcommand\T{\rule{0pt}{2.6ex}}
\newcommand\B{\rule[-1.2ex]{0pt}{0pt}}
  \begin{center}
    \caption{Summary of systematic uncertainties in the expected numbers of events for
     a dilepton mass of 1~\TeV\ ~(2~\TeV).
     NA indicates that the uncertainty is not applicable.
    \label{tab:systematics}}
    \centering
    \vspace{0.2cm}
    \begin{tabular}{lcccc}
      \hline
      \hline
      Source \T             & \multicolumn{2}{c}{$e e$} & \multicolumn{2}{c}{$\mu \mu$} \\
             \B             & Signal        & Background      & Signal        & Background \\
      \hline
      Normalization \T      & 5\% ~(5\%)     & NA               & 5\% ~(5\%)     & NA \\
      PDFs/$\alpha_{\rm S}$/scale & NA             & 7\% ~(20\%)      & NA            & 7\% ~(20\%) \\
      Electroweak $K$ factor & NA             & 2.3\% ~(4.5\%)   & NA            & 2.3\% ~(4.5\%) \\
      Efficiency            & 1.0\% ~(2.0\%) & 1.0\% ~(2.0\%)   & 3.0\% ~(6.0\%) & 3.0\% ~(6.0\%) \\
      Scale/Resolution      & 1.2\% ~(2.4\%) & 1.2\% ~(2.4\%)   & 1.2\% ~(12\%) & 1.2\% ~(12\%) \\
      Multijet/$W$+jets background \B     & NA             & 12\% ~(26\%)     & NA            & $< 0.1$\% \\
      \hline
      Total  \T \B          & 5\% ~(6\%)     & 14\%  ~(33\%)    & 6\% ~(14\%)   & 8\% ~(25\%) \\
      \hline
      \hline
    \end{tabular}
  \end{center}
\end{table*}

\section{\label{sec:analysis}Statistical Analysis}
The data analysis proceeds with a Bayesian method to compare the observed
event yields with the expected yields for a range of different NP
model parameters (where the NP corresponds to either contact interactions or
large extra dimensions).
Specifically, the number of expected events in a given search region
is 
\begin{equation}
\mu = n_{\rm DY+NP}(\theta,\bar{\nu}) + n_{\rm non-DY~bg}(\bar{\nu})~,
\end{equation}
where $n_{\rm DY+NP}(\theta,\bar{\nu})$ is the number of events predicted
by the DY+NP simulation for a particular choice of NP model 
parameter $\theta$, $n_{\rm non-DY~bg}(\bar{\nu})$ is the number of non-DY background
events,
and $\bar{\nu}$ represents the set of Gaussian nuisance parameters that account
for systematic uncertainties.
The parameter $\theta$ corresponds to a choice of energy scale $\Lambda$ and
interference parameter $\eta_{\rm LL}$ in the CI analysis or to a choice of
string scale $M_{\rm S}$ and formalism in the ADD analysis.
In the case of the CI analysis, the input to evaluate the 
complete set of $\mu$ values is shown 
in Tables~\ref{yields_elec} and \ref{yields_muon} 
for the dielectron and dimuon channels, respectively.
For each mass bin, a second-order polynomial is used to model the dependence
of $\mu$ on $1/\Lambda^2$.
In the case of the ADD analysis, $\mu$ is also parameterized by a 
second-order polynomial but as a function of $1/M_{\rm S}^4$.

The likelihood of observing a set of $\bar{n}$ events in $N$ invariant mass bins
is given by a product of Poisson probabilities for each mass bin $k$:
\begin{equation}\label{likelihood}
 \text{$\cal L$} (\bar{n} \mid \theta,\bar{\nu}) = \displaystyle\prod_{k=1}^{N}\frac{\mu_k^{n_k}e^{-\mu_k}}{n_k!}.
\end{equation}
According to Bayes' theorem, the posterior probability
for the parameter $\theta$ given $\bar{n}$ observed events
is
\begin{equation}\label{posterior}
 \text{$\cal P$} (\theta \mid \bar{n}) = \frac{1}{\text{$\cal Z$}} \text{$\cal L_{\rm M}$} (\bar{n} \mid \theta) P(\theta),
\end{equation}
where ${\cal Z}$ is a normalization constant and
the marginalized likelihood $\cal L_{\rm M}$ corresponds to the likelihood after
all nuisance parameters have been integrated out.
This integration is performed assuming that the nuisance parameters are correlated
across all mass bins; Table~\ref{tab:systematics} shows which parameters are taken into
account for either or both of the signal and background expectations.
The prior probability $P(\theta)$ is chosen to be flat in either $1/\Lambda^2$ or
$1/\Lambda^4$ for the CI analysis, and either
$1/M_{\rm S}^4$ or $1/M_{\rm S}^8$ for the ADD analysis.
These choices are motivated by the form of Eqs.~(\ref{cross_section_CI})
and (\ref{cross_section_ADD}).
The 95\% C.L. limit is then obtained by finding the value 
$\theta_{\rm lim}$ satisfying
$\int_0^{\theta_{\rm lim}} \text{$\cal P$} (\theta \mid \bar{n}) \: d\theta = 0.95$,
where $\theta$ is chosen to be  $1/\Lambda^2$, $1/\Lambda^4$, $1/M_{\rm S}^4$ or $1/M_{\rm S}^8$.
The above calculations have been performed with the Bayesian Analysis Toolkit (BAT)~\cite{Caldwell:2008fw},
which uses a Markov chain Monte Carlo technique to integrate over
nuisance parameters.

\section{\label{sec:CIresults}Contact Interactions Analysis and Results}

To test the consistency between the data and the SM
in the CI search region ($m_{\ell\ell} > 400$~\GeV),
a likelihood ratio test is performed by producing a set of
SM-like pseudoexperiments and comparing the likelihood ratio
between the signal+background and pure background hypotheses obtained
in the data to the results of the pseudoexperiments.
The signal+background likelihood is evaluated 
at the $\Lambda$ value that maximizes it.
The derived {\it p}-value, corresponding to the probability of observing
a fluctuation in the pseudoexperiments
that is at least as signal-like as that seen in the data
(i.e., with a maximum likelihood ratio greater than or equal to that obtained in the data),
is estimated to be 15\% (76\%) in the dielectron channel
and 79\% (59\%) in the dimuon channel for constructive (destructive) interference.
These values indicate that there is no significant evidence for
contact interactions in the analyzed data and thus limits are
set on the contact interaction scale $\Lambda$.

Limits are obtained with the Bayesian method described above.
Electroweak corrections are applied
to both DY and DY+CI samples for consistency, 
although part of the electroweak corrections cannot
be computed reliably due to the unknown new phenomena represented by the
contact interaction. This particular choice results in slightly more conservative limits. 

The expected 95\% C.L. lower limit values on the energy scale $\Lambda$ are found to be
$\expLimitMinusElec \pm 1.7$ \TeV{} for constructive interference ($\Lambda^-$)
and $\expLimitPlusElec~\pm 1.0$ \TeV{} for destructive interference ($\Lambda^+$)
in the dielectron channel.
The corresponding expected limits in the dimuon channel are
$\expLimitMinusMuon \pm 1.5$ \TeV{} and $\expLimitPlusMuon \pm 1.1$ \TeV{}.
The quoted uncertainties correspond to the 68\% range of limits surrounding the
median value (taken to be the expected limit) of all limits obtained with a set of 
pseudoexperiments.
Limits are expected to be stronger in the dielectron channel than in the dimuon channel
due to the significantly larger acceptance for the dielectron selection.

The observed limits (at 95\% C.L.) are
$\Lambda^- > \obsLimitMinusElec$~\TeV{} and $\Lambda^+ > \obsLimitPlusElec$~\TeV{}
in the dielectron channel for constructive and destructive interference, respectively.
The corresponding limits in the dimuon channel are
$\Lambda^- > \obsLimitMinusMuon$~\TeV{} and $\Lambda^+ > \obsLimitPlusMuon$~\TeV{}.
These limits are summarized in Table~\ref{limits_CI}.

If instead of choosing the prior to be flat in $1/\Lambda^2$,
it is selected to be flat in $1/\Lambda^4$ to match the form of the
pure CI term in Eq.~(\ref{cross_section_CI}), the observed limit in the
dielectron channel becomes weaker
by 0.7~\TeV{} for constructive interference
and  0.4~\TeV{} for destructive interference. 
The corresponding respective shifts to lower values are 1.2 and 0.6~\TeV{} 
in the dimuon channel, see Table~\ref{limits_CI}.

Finally, a limit is set for the combination of the dielectron and dimuon channels, 
assuming lepton universality, by computing a combined posterior probability 
for the two channels. The following sources of systematic
uncertainty are treated as fully correlated between the two channels: 
PDF and $\alpha_{\rm S}$, QCD and electroweak $K$ factors,
and \Zgstr\ cross section for normalization.
All other sources are treated as uncorrelated.
The resulting combined limits are
$\Lambda^- > \obsLimitMinusComb$~\TeV{} and $\Lambda^+ > \obsLimitPlusComb$~\TeV{}
for the $1/\Lambda^2$ prior.
Table~\ref{limits_CI} summarizes all limits for the two priors considered in
this analysis.

\begin{table}[!tb]
\newcommand\T{\rule{0pt}{2.6ex}}
\newcommand\B{\rule[-1.2ex]{0pt}{0pt}}
  \begin{center}
    \caption{Expected and observed 95\% C.L. lower limits on the contact interaction
     energy scale $\Lambda$ for the dielectron and dimuon channels, as well as for
     the combination of those channels. Results are provided for constructive and
     destructive interference as well as
     different choices of flat priors: $1/\Lambda^2$ and $1/\Lambda^4$.\label{limits_CI}}
    \begin{tabular}{lccccc}
      \hline\hline
      Channel \T & Prior & \multicolumn{2}{c}{Expected limit [\TeV{}]} & \multicolumn{2}{c}{Observed limit [\TeV{}]} \\
              &       & Constr. & Destr. & Constr. & Destr. \\
      \hline
      $e e$ \T & $1/\Lambda^2$ & $\expLimitMinusElec$ & $\expLimitPlusElec$ & $\obsLimitMinusElec$ & $\obsLimitPlusElec$ \\
              \B  & $1/\Lambda^4$ & 12.5 & 9.8 & 11.4 & 9.1 \\ \hline
      $\mu \mu$ \T & $1/\Lambda^2$ & $\expLimitMinusMuon$ & $\expLimitPlusMuon$ & $\obsLimitMinusMuon$ & $\obsLimitPlusMuon$ \\
              \B  & $1/\Lambda^4$ & 11.6 & 9.1 & 11.7 & 9.0 \\ \hline
      $ee + \mu\mu$ \T & $1/\Lambda^2$ & $\expLimitMinusComb$ & $\expLimitPlusComb$ & $\obsLimitMinusComb$ & $\obsLimitPlusComb$ \\
              \B & $1/\Lambda^4$ & 13.8 & 10.5 & 12.9 & 9.8 \\
      \hline\hline
    \end{tabular}
  \end{center}
\end{table}

\section{\label{sec:ADDresults}Large Extra Dimensions Analysis and Results}

The search for large extra dimensions is carried out similarly to that for contact
interactions.
A difference from the CI analysis is that the DY component present in the
{\sc Sherpa} DY+ADD simulated samples is subtracted out to compute the net ADD contribution
to the total event yield. The DY background is modeled with the same {\sc Pythia}
DY sample as is used for the CI analysis.
Another difference is that the search is performed in only one mass bin with
the minimum mass chosen at the value giving the strongest expected limit. 
This optimization results in a signal region with a minimum mass requirement 
of 1300 \GeV{} as determined from a set of pseudoexperiments in each of 
the dielectron and dimuon channels.
Table~\ref{yields_ADD} presents the expected and observed event yields in the signal region,
including the expectation for several $M_{\rm S}$ values in the GRW formalism.
\begin{table}[!tb]
\newcommand\T{\rule{0pt}{2.6ex}}
\newcommand\B{\rule[-1.2ex]{0pt}{0pt}}
  \begin{center}
    \caption{Expected and observed number of events with $m_{\ell\ell} >$ 1300 \GeV{}
in the dielectron and dimuon channels. 
Yields given for different $M_{\rm S}$ values correspond to the sum of signal and background events, with the signal obtained in the GRW formalism.
All yields are normalized to the $Z$ peak control region.
The errors quoted originate from systematic uncertainties and limited MC statistics.}
    \label{yields_ADD}
    \vspace{0.2cm}
    \begin{tabular}{lcc}
     \hline\hline
      Process \T \B  & $ee$           & $\mu\mu$  \\ \hline
      DY  \T         & $0.89   \pm 0.21$   & $0.54 \pm 0.16$ \\
      $\ttbar$       &  $<0.01$            & $<0.01$  \\
      Diboson        & $0.075  \pm 0.005$  & $0.059 \pm 0.010$  \\
      Multijet/$W$+jets \B & $0.16  \pm 0.20$  & --  \\
     \hline
     Total background \T \B     & $1.13  \pm 0.29$  & $0.60 \pm 0.16$  \\
     \hline
     $M_{\rm S} = 1.5$ \TeV \T & $72   \pm   5$   & $47   \pm 9$   \\
     $M_{\rm S} = 2.0$ \TeV    & $40.2 \pm 2.6$   & $22   \pm 4$ \\
     $M_{\rm S} = 2.5$ \TeV    & $11.7 \pm 0.9$   & $6.3  \pm 1.1$ \\
     $M_{\rm S} = 3.0$ \TeV \B & $4.2  \pm 0.4$   & $2.3  \pm 0.4$ \\
     \hline
     Data  \T \B    & 2                 & 0 \\
     \hline\hline
    \end{tabular}
  \end{center}
\end{table}

  The consistency between the number of observed events in the data and the
predicted SM contribution is assessed using a set of SM-like pseudoexperiments.
Using the same likelihood ratio approach as for the CI analysis,
{\it p}-values of 6\% and 68\% are obtained
in the dielectron and dimuon channels, respectively.
These values indicate that there is no significant evidence for large extra dimensions
and thus limits are set on the scale $M_{\rm S}$.
The observed limits are $M_{\rm S} > $ 2.73 (2.62)~\TeV{} in the dielectron channel
and $M_{\rm S} > $ 2.83 (2.61)~\TeV{} in the dimuon channel at 95\% C.L. with
a prior flat in 1/$M_{\rm S}^{4}$ (1/$M_{\rm S}^{8}$). Table~\ref{limits_ADD} shows
these observed limits along with the expected limits. Limits in the
dielectron channel are slightly worse than expected due to the larger number
of events observed in the data compared with the SM expectation.
The dielectron and dimuon channels are combined taking correlated systematic uncertainties 
into account in a way identical to the CI analysis.

\begin{table}[!tb]
\newcommand\T{\rule{0pt}{2.6ex}}
\newcommand\B{\rule[-1.2ex]{0pt}{0pt}}
  \begin{center}
    \caption{Expected and observed 95\% C.L. lower limits on $M_{\rm S}$ in the dielectron
     and dimuon channels, as well as for the combination of those channels without
     and with the diphoton channel in the GRW formalism.
     Separate results are provided for the different choices of flat priors:
     $1/M_{\rm S}^4$ and $1/M_{\rm S}^8$.\label{limits_ADD}}
    \begin{tabular}{lccc}
      \hline\hline
      Channel \T \B & Prior & \multicolumn{1}{c}{Exp. limit [\TeV{}]} & \multicolumn{1}{c}{Obs. limit [\TeV{}]} \\
      \hline
      $e e$ \T                        & $1/M_{\rm S}^4$ & 2.88 & 2.73 \\
            \B                        & $1/M_{\rm S}^8$ & 2.72 & 2.62 \\ \hline
      $\mu \mu$ \T                    & $1/M_{\rm S}^4$ & 2.83 & 2.83 \\
                \B                    & $1/M_{\rm S}^8$ & 2.61 & 2.61 \\ \hline
      $ee + \mu\mu$ \T                & $1/M_{\rm S}^4$ & 3.16 & 3.00 \\
                    \B                & $1/M_{\rm S}^8$ & 2.96 & 2.85 \\ \hline
      $ee + \mu\mu + \gamma\gamma$ \T & $1/M_{\rm S}^4$ & 3.43 & 3.22 \\
                                   \B & $1/M_{\rm S}^8$ & 3.27 & 3.12 \\
      \hline\hline
    \end{tabular}
  \end{center}
\end{table}

  A search for large extra dimensions has also been carried out in the diphoton final
state using the data sample collected by ATLAS in 2011~\cite{Aad:2012cy}.
The results of that search are combined with the dilepton results presented here
with the use of BAT. Correlated sources of systematic uncertainty 
are treated as follows. The PDF uncertainty in the SM diphoton and DY production is
considered to be fully correlated between the $ee$, $\mu\mu$ and $\gamma\gamma$ channels,
whereas the multijet background uncertainty is fully correlated between the $ee$ and $\gamma\gamma$
final states.
It should be noted that the $ee$ and $\gamma\gamma$ samples are statistically
uncorrelated since $ee$ candidates have been explicitly removed from the diphoton
analysis at the event selection stage.
The observed and expected combined limits are given in Table~\ref{limits_ADD}, with
the most stringent observed limit obtained for the dilepton-diphoton combination:
$M_{\rm S} >$ 3.22 (3.12) \TeV\ with a prior flat in $1/M_{\rm S}^4$ ($1/M_{\rm S}^8$)
in the GRW formalism.

The limits obtained using the GRW formalism have been translated into the Hewett and
HLZ formalisms using Eq.~(\ref{ADDformalism}) with results shown in 
Table~\ref{limits_ADD_formalisms}.
Limits are also obtained with a $K$ factor applied to the ADD signal yield 
to account for next-to-leading-order QCD corrections.
A constant $K$ factor of 1.6 is applied in the dilepton 
channel~\cite{*[{}] [{, including updated calculations at this collision energy.}]
 Kumar:2006id,*Mathews:2004xp}
and 1.7 in the diphoton channel~\cite{Kumar:2008pk,*Kumar:2009nn}.
The dilepton-diphoton combination increases limits by approximately 0.2 (0.3) \TeV{}
with a prior flat in $1/M_{\rm S}^4$ ($1/M_{\rm S}^8$), taking
QCD corrections into account as shown in Table~\ref{limits_ADD_formalisms_kf}.

\begin{table}[!tb]
\newcommand\T{\rule{0pt}{2.6ex}}
\newcommand\B{\rule[-1.2ex]{0pt}{0pt}}
  \begin{center}
    \caption{Observed 95\% C.L. lower limits on $M_{\rm S}$ (in units of \TeV{}),
     including systematic uncertainties,
     for ADD signal in the GRW, Hewett and HLZ formalisms with no $K$ factor applied to
     the signal.
     Separate results are provided for the different choices of flat priors:
     $1/M_{\rm S}^4$ and $1/M_{\rm S}^8$.
      \label{limits_ADD_formalisms}}
    \vspace{0.2cm}
\begin{tabular}{lcccccccc}
\hline\hline
Channel \T & Prior & GRW & Hewett & \multicolumn{5}{c}{HLZ} \\
           &       &     &        & $n$=3 & $n$=4 & $n$=5 & $n$=6 & $n$=7 \\ \hline
$ee$ \T \B     & $1/M_{\rm S}^{4}$ & 2.73 & 2.44 & 3.25 & 2.73 & 2.47 & 2.30 & 2.17 \\
        \B     & $1/M_{\rm S}^{8}$ & 2.62 & 2.48 & 2.86 & 2.62 & 2.49 & 2.40 & 2.34 \\ \hline
$\mu\mu$ \T \B   & $1/M_{\rm S}^{4}$ & 2.83 & 2.52 & 3.36 & 2.83 & 2.55 & 2.38 & 2.25 \\
            \B   & $1/M_{\rm S}^{8}$ & 2.61 & 2.47 & 2.85 & 2.61 & 2.48 & 2.40 & 2.33 \\ \hline
$ee$ + $\mu\mu$ \T \B   & $1/M_{\rm S}^{4}$ & 3.00 & 2.68 & 3.57 & 3.00 & 2.71 & 2.52 & 2.39 \\
                   \B   & $1/M_{\rm S}^{8}$ & 2.85 & 2.70 & 3.11 & 2.85 & 2.71 & 2.62 & 2.54 \\ \hline
$ee + \mu\mu$ \T \B & $1/M_{\rm S}^{4}$ & 3.22 & 2.88 & 3.83 & 3.22 & 2.91 & 2.71 & 2.56 \\
$~ + \gamma\gamma$ \B & $1/M_{\rm S}^{8}$ & 3.12 & 2.95 & 3.40 & 3.12 & 2.96 & 2.86 & 2.78 \\ \hline\hline
\end{tabular}
  \end{center}
\end{table}

\begin{table}[!tb]
\newcommand\T{\rule{0pt}{2.6ex}}
\newcommand\B{\rule[-1.2ex]{0pt}{0pt}}
  \begin{center}
    \caption{Observed 95\% C.L. lower limits on $M_{\rm S}$ (in units of \TeV{}), 
     including systematic uncertainties,
     for ADD signal in the GRW, Hewett and HLZ formalisms with $K$ factors of 1.6 and 1.7 applied to
     the signal for the dilepton and diphoton channels, respectively.
     Separate results are provided for the different choices of flat priors:
     $1/M_{\rm S}^4$ and $1/M_{\rm S}^8$.
      \label{limits_ADD_formalisms_kf}}
    \vspace{0.2cm}
\begin{tabular}{lcccccccc}
\hline\hline
Channel \T & Prior & GRW & Hewett   & \multicolumn{5}{c}{HLZ} \\
           &       &     &          &  $n$=3 & $n$=4 & $n$=5 & $n$=6 & $n$=7 \\ \hline
$ee$ \T    & $1/M_{\rm S}^{4}$ & 2.95 & 2.63 & 3.51 & 2.95 & 2.66 & 2.48 & 2.34 \\
     \T \B & $1/M_{\rm S}^{8}$ & 2.82 & 2.67 & 3.08 & 2.82 & 2.68 & 2.59 & 2.52 \\ \hline
$\mu\mu$ \T    & $1/M_{\rm S}^{4}$ & 3.07 & 2.74 & 3.65 & 3.07 & 2.77 & 2.58 & 2.44 \\
         \T \B & $1/M_{\rm S}^{8}$ & 2.82 & 2.67 & 3.08 & 2.82 & 2.68 & 2.59 & 2.52 \\ \hline
$ee$ + $\mu\mu$ \T    & $1/M_{\rm S}^{4}$ & 3.27 & 2.92 & 3.88 & 3.27 & 2.95 & 2.75 & 2.60 \\
                \T \B & $1/M_{\rm S}^{8}$ & 3.09 & 2.92 & 3.37 & 3.09 & 2.94 & 2.84 & 2.76 \\ \hline
$ee + \mu\mu$ \T  & $1/M_{\rm S}^{4}$ & 3.51 & 3.14 & 4.18 & 3.51 & 3.17 & 2.95 & 2.79 \\
$~ + \gamma\gamma$ \T \B & $1/M_{\rm S}^{8}$ & 3.39 & 3.20 & 3.69 & 3.39 & 3.22 & 3.11 & 3.02 \\ \hline\hline
\end{tabular}
  \end{center}
\end{table}

\section{\label{sec:conclusions}Conclusions}
A search for contact interactions and large extra dimensions
has been performed in dielectron and dimuon events
produced in LHC proton-proton collisions at $\sqrt{s} = 7$~\TeV{}.
The data sample corresponds to an integrated luminosity of
$\integLumiElec$ ($\integLumiMuon$) fb$^{-1}$ of $pp$ collisions
in the dielectron (dimuon) channel recorded with the ATLAS detector.
No significant deviation from the Standard Model is observed in
the dilepton mass distributions.
Using a Bayesian approach with a prior flat in $1/\Lambda^2$, as was
done in most previous searches at hadron colliders,
the following 95\% C.L. limits are set on the energy scale of contact interactions:
$\Lambda^- > \obsLimitMinusElec$~\TeV{} ($\Lambda^+ > \obsLimitPlusElec$~\TeV{})
in the dielectron channel and
$\Lambda^- > \obsLimitMinusMuon$~\TeV{} ($\Lambda^+ > \obsLimitPlusMuon$~\TeV{})
in the dimuon channel
for constructive (destructive) interference in the left-left isoscalar
compositeness model.
Somewhat weaker limits are obtained with a prior flat in $1/\Lambda^4$.
These limits improve existing bounds on  $e e q q$ and $\mu\mu qq$ contact interactions
from a single experiment.
Limits are also set on the scale $M_{\rm S}$ in the ADD large extra dimensions model.
Those range from 2.4 to 3.9 \TeV\ depending on the choice of model, channel, and prior.
After combining the dilepton and diphoton searches, the limits are
in the range from 2.6 to 4.2 \TeV{}.

~~~

% Acknowledgements for papers with collision data
% Version 24-Sep-2012

\section*{Acknowledgements}

We thank CERN for the very successful operation of the LHC, as well as the
support staff from our institutions without whom ATLAS could not be
operated efficiently.

We acknowledge the support of ANPCyT, Argentina; YerPhI, Armenia; ARC,
Australia; BMWF and FWF, Austria; ANAS, Azerbaijan; SSTC, Belarus; CNPq and FAPESP,
Brazil; NSERC, NRC and CFI, Canada; CERN; CONICYT, Chile; CAS, MOST and NSFC,
China; COLCIENCIAS, Colombia; MSMT CR, MPO CR and VSC CR, Czech Republic;
DNRF, DNSRC and Lundbeck Foundation, Denmark; EPLANET and ERC, European Union;
IN2P3-CNRS, CEA-DSM/IRFU, France; GNSF, Georgia; BMBF, DFG, HGF, MPG and AvH
Foundation, Germany; GSRT, Greece; ISF, MINERVA, GIF, DIP and Benoziyo Center,
Israel; INFN, Italy; MEXT and JSPS, Japan; CNRST, Morocco; FOM and NWO,
Netherlands; BRF and RCN, Norway; MNiSW, Poland; GRICES and FCT, Portugal; MERYS
(MECTS), Romania; MES of Russia and ROSATOM, Russian Federation; JINR; MSTD,
Serbia; MSSR, Slovakia; ARRS and MVZT, Slovenia; DST/NRF, South Africa;
MICINN, Spain; SRC and Wallenberg Foundation, Sweden; SER, SNSF and Cantons of
Bern and Geneva, Switzerland; NSC, Taiwan; TAEK, Turkey; STFC, the Royal
Society and Leverhulme Trust, United Kingdom; DOE and NSF, United States of
America.

The crucial computing support from all WLCG partners is acknowledged
gratefully, in particular from CERN and the ATLAS Tier-1 facilities at
TRIUMF (Canada), NDGF (Denmark, Norway, Sweden), CC-IN2P3 (France),
KIT/GridKA (Germany), INFN-CNAF (Italy), NL-T1 (Netherlands), PIC (Spain),
ASGC (Taiwan), RAL (UK) and BNL (USA) and in the Tier-2 facilities
worldwide.

%%%%%%%%%%%%%%%%%%%%%%%%%%%%%%%%%%%%%%%%%%%%%%%%%%%%%%%%%%%%%%%%%%%%%%%%%%%%%%%
% Bibliography
%%%%%%%%%%%%%%%%%%%%%%%%%%%%%%%%%%%%%%%%%%%%%%%%%%%%%%%%%%%%%%%%%%%%%%%%%%%%%%%

%\bibliographystyle{plainnat}
%\bibliographystyle{atlasnote}
\bibliographystyle{atlasBibStyleWoTitle}
\bibliography{cied_dilepton_7tev}

\onecolumngrid
\clearpage
\input{atlas_authlist}

\end{document}
%
% ****** End of file template.aps ******

%% file: atlas_authlist.tex
% ATLAS Collaboration author list
% Data extracted on 30-Nov-2012 for paper reference EXOT-2012-17
%\documentclass[11pt]{article}
%\usepackage{a4wide}\begin{document}
\begin{flushleft}
{\Large The ATLAS Collaboration}

\bigskip

G.~Aad$^{\rm 48}$,
T.~Abajyan$^{\rm 21}$,
B.~Abbott$^{\rm 111}$,
J.~Abdallah$^{\rm 12}$,
S.~Abdel~Khalek$^{\rm 115}$,
A.A.~Abdelalim$^{\rm 49}$,
O.~Abdinov$^{\rm 11}$,
R.~Aben$^{\rm 105}$,
B.~Abi$^{\rm 112}$,
M.~Abolins$^{\rm 88}$
% ...
\bigskip
\\
% ...
$^{11}$ Institute of Physics, Azerbaijan Academy of Sciences, Baku, Azerbaijan\\
$^{12}$ Institut de F{\'\i}sica d'Altes Energies and Departament de F{\'\i}sica de la Universitat Aut{\`o}noma de Barcelona and ICREA, Barcelona, Spain\\
% ...
$^{21}$ Physikalisches Institut, University of Bonn, Bonn, Germany\\
% ...
$^{48}$ Fakult{\"a}t f{\"u}r Mathematik und Physik, Albert-Ludwigs-Universit{\"a}t, Freiburg, Germany\\
$^{49}$ Section de Physique, Universit{\'e} de Gen{\`e}ve, Geneva, Switzerland\\
% ...
$^{88}$ Department of Physics and Astronomy, Michigan State University, East Lansing MI, United States of America\\
% ...
$^{105}$ Nikhef National Institute for Subatomic Physics and University of Amsterdam, Amsterdam, Netherlands\\
% ...
$^{111}$ Homer L. Dodge Department of Physics and Astronomy, University of Oklahoma, Norman OK, United States of America\\
$^{112}$ Department of Physics, Oklahoma State University, Stillwater OK, United States of America\\
% ...
$^{115}$ LAL, Universit{\'e} Paris-Sud and CNRS/IN2P3, Orsay, France\\
% ...
\end{flushleft}

%\end{document}
% Created with /afs/cern.ch/user/a/atlaspo/svntools/Upload/xml2latex.py